\documentclass[11pt,a4]{article}
\usepackage[T1]{fontenc}
\usepackage{textcomp}
\usepackage{graphicx}
\usepackage{amsmath,amssymb}
\usepackage{mathtools}
\usepackage[labelfont=bf]{caption,subcaption}
\usepackage{float}
\usepackage[margin=2.5cm]{geometry}
\usepackage{authblk}
\usepackage{xcolor}
\definecolor{ceruleanblue}{rgb}{0.16, 0.32, 0.75}
\usepackage[colorlinks=true,allcolors=ceruleanblue]{hyperref}
\usepackage{siunitx}

\usepackage[backend=biber,
style=phys,
biblabel=brackets,
bibencoding=utf8]{biblatex}
\usepackage[normalem]{ulem}
\usepackage{parskip}
\def\pa{\partial}

\def\ii{\textrm i}

\newcommand{\be}{\begin{equation}}
\newcommand{\en}{\end{equation}}
\newcommand{\bi}{\begin{itemize}}
\newcommand{\ei}{\end{itemize}}

\date{}
\newcommand*{\tran}{{\mkern-1.5mu\mathsf{\footnotesize{T}}}}
\let\Re\relax
\let\Im\relax
\DeclareMathOperator{\Re}{\mathrm{Re}}
\DeclareMathOperator{\Im}{\mathrm{Im}}

\begin{document}

\title{Zig-zag dynamics in a Stern-Gerlach spin measurement}

\author[*,$\ddagger$]{Simon Krekels}
\author[*]{Christian Maes}
\author[*]{Kasper Meerts}
\author[*,$\dagger$]{Ward Struyve}
\affil[*]{Department of Physics and Astronomy, KU Leuven}
\affil[$\dagger$]{Centre for Logic and Philosophy of Science, KU 
Leuven}
\affil[$\ddagger$]{Imec, Leuven}

\maketitle
\begin{abstract}
The one-century-old Stern-Gerlach setup is paradigmatic for a quantum measurement. We visualize the electron trajectories following the Bohmian zig-zag dynamics. This dynamics was developed in order to deal with the fundamentally massless nature of particles (with mass emerging from the Brout-Englert-Higgs mechanism). The corresponding trajectories exhibit a stochastic zig-zagging, as the result of the coupling between left- and right-handed chiral Weyl states. This zig-zagging persists in the nonrelativistic limit, which will be considered here, and which is described by the Pauli equation for a nonuniform external magnetic field. Our results clarify the different meanings of ``spin'' as a property of the wave function and as a random variable in the Stern-Gerlach setup, and they illustrate the notion of effective collapse. We also examine the case of an EPR-pair.  By letting one of the entangled particles pass through a Stern-Gerlach device, the nonlocal influence (action-at-a-distance) on the other particle is manifest in its trajectory, {\it e.g.} by initiating its zig-zagging.
\end{abstract}

\textit{Dedicated to the memory of Detlef Dürr, friend and mentor.}

\section{Introduction} 
Bohmian mechanics (also known as the de\ Broglie-Bohm or pilot-wave theory) describes the motion of point-particles guided by the wave function \cite{bohm93,holland93b,duerr09}.
The nonrelativistic dynamics originally presented by de\ Broglie and Bohm is deterministic. The extension of this dynamics to quantum field theory proposed by Dürr \textit{et al}.\ introduces a stochastic element \cite{duerr02,duerr032,duerr04b,duerr05a}. More specifically, the events of particle creation and annihilation are stochastic, interrupting the otherwise deterministic motion. 
Furthermore, taking the particles to be fundamentally massless, as suggested by the Standard Model of particle physics, implies that even single fermions will evolve stochastically, with the motion alternatingly being determined by a left- and right-handed chiral Weyl spinor \cite{colin11,struyve12}.{\footnote{This dynamics was also presented before for a 2-dimensional space-time \cite{deangelis86}.}} The resulting particle dynamics exhibits a zig-zag like motion; the trajectories are continuous, but the velocities are not.\footnote{The Feynman diagrams display a similar zig-zag behavior \cite[p. 630]{penrose04}.}
This type of motion is analogous to the run-and-tumble dynamics, which is well known for the description of, e.g., self-propelling bacteria \cite{schnitzer1993,patteson2015}. 

For quite some time, the Bohmian trajectories have been studied and plotted in a variety of situations using the original deterministic Bohmian dynamics.
The zig-zag dynamics on the other hand, has so far received little attention.
In a previous study, we considered the nonrelativistic limit of this dynamics in the case of the double-slit experiment \cite{maes22}.
Here, we continue that study in the context of another paradigmatic quantum measurement: the now centennial Stern-Gerlach experiment.

There are several reasons to be interested in visualizing these trajectories.
To start, there is the important fact that this is possible at all, despite earlier beliefs that a trajectory description of quantum systems must be fundamentally flawed. Secondly, it is helpful, much in the same sense that a trajectory-based description of Brownian motion through the Langevin equation is an essential methodological complement to its Fokker-Planck equation, which describes the evolution of densities.
Furthermore, there are advantages related to specific questions and illustrations; for example, the making explicit of nonlocality, a fundamental feature of nature as shown by Bell, building on the work of Einstein, Podolsky and Rosen \cite{bell}.
Or, as another example, the visualization of the formation of the interference pattern in double-slit experiments \cite{philippidis79,holland03a,gondran05,maes22}.

\begin{figure}
    \centering
    \includegraphics[width=0.7\textwidth]{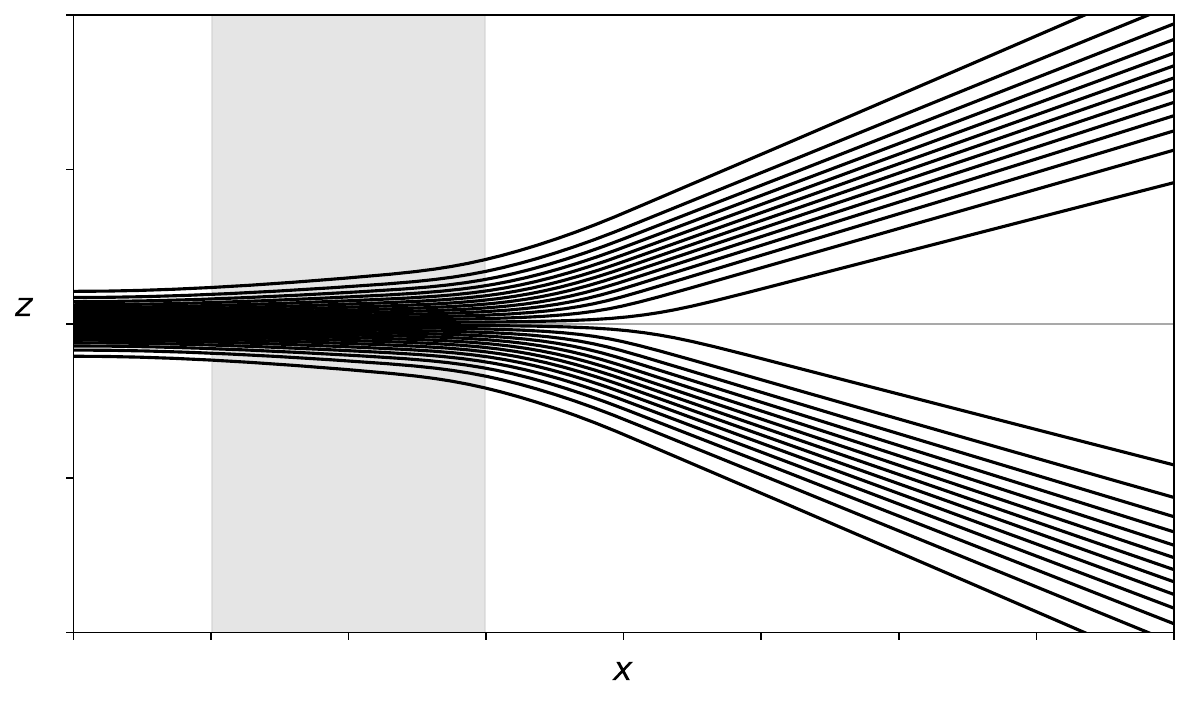}
    \caption{ 
    Some Bohmian trajectories for a Stern-Gerlach spin measurement in the $z$-direction, where the initial spinor has spin {\em up} in the $y$-direction (with the $y$-axis pointing into the plane of the paper, not drawn). The grey area indicates where the magnetic field is present.}
    \label{fig:sg_standard}
\end{figure}

In the present work, we concentrate on \emph{measuring spin} which has its own history of confusion.
While the Stern-Gerlach experiment is sometimes described as \emph{the discovery of spin},  its initial history is more complicated.
The original experiment of 1922 was motivated by disbelief on the part of Stern in the Bohr-Sommerfeld quantization of the direction of the angular momentum. 
\begin{quote}
   If this nonsense of Bohr should in the end prove to be right, we will quit physics! \\
--- Oath sworn by Otto Stern and Max von Laue.
 \end{quote}
In other words, Stern wanted to test the hypothesis of directional quantization, as explicitly stated by Bohr (1913) in his atomic model, and as extended by Sommerfeld (1916).
He found Gerlach ready for a ``magnetic'' experiment, and their experimental setup has been well-known ever since.
In the original experiment, a narrow beam of silver atoms was shot through a region with a nonuniform magnetic field.
Upon leaving the magnetic field, the atoms hit a detection screen and the beam was observed to be split in two.
The classical prediction would be a continuous distribution, as for randomly oriented magnetic dipoles.
Pauli wrote to Gerlach, ``This should convert even the nonbeliever Stern.''. 
The Stern-Gerlach publication \cite{sterngerlach1922richtings} was given the title ``Experimental evidence of directional quantization in a magnetic field.''

However, as a matter of fact, silver has no orbital angular momentum; what Stern and Gerlach measured was \emph{spin}, an intrinsic form of angular momentum.
It was Kronig in 1925 who proposed electron spin, yet never published his proposal.
After finding the idea ``quite ridiculous'', Pauli wrote about ``a classically non-describable two-valuedness'' \cite{pauli1925}.
A few months later, Uhlenbeck and Goudsmit used these ideas to convincingly explain the Zeeman effect \cite{goudsmit_uhlenbeck1925,uhlenbeck1926}, and in that way, they were the first to explicitly introduce electron spin.
See \cite{uhl,friedrich03,friedrich21,schmidt16,maes2023} for more history and references. 

The main aim of the present paper is to visualize the influence of spin (coupled to an inhomogeneous magnetic field) on the motion of electrons in a Stern-Gerlach-like setup. That problem has been treated before with the standard Bohmian dynamics \cite{dewdney86,dewdney87a,dewdney87b,dewdney88b,gondran14,norsen14b}, to generate trajectories as displayed in Fig.~\ref{fig:sg_standard}. Here, we consider the zig-zag dynamics in the nonrelativistic limit, with the Pauli equation as governing wave equation.

The original Stern-Gerlach experiment did not involve free electrons, but silver atoms. The electric neutrality of the atoms allows the Lorentz force to be ignored, so that the only interaction is the spin-field coupling. For electrons, on the other hand, it has been argued that the usual Stern-Gerlach setup cannot be used because the Lorentz force would blur the splitting of the beams \cite{mott29,pauli32}. Alternative setups have been proposed, however, which could result in a spin separation of electron beams, see e.g.\ \cite{batelaan97,garraway99,mcgregor15}. In our work, we make use of a toy interaction, modeled after the usual Stern-Gerlach setup, which serves well to illustrate the zig-zag dynamics, but whose physical implementation is left open.

Visualizing the dynamics also allows us to  distinguish the various notions of ``spin'' that are at play. That is, we say that the electron has spin 1/2, meaning that its wave function is a spinor. Another thing is to \emph{measure spin} in a particular direction, with outcomes that can be \emph{up} or \emph{down}. The word ``measurement'' then suggests that the experiment reveals some pre-determined value for the spin, but this is not the case. Instead, just as in the case of the original Bohmian dynamics \cite{duerr04,bohm93,holland93b,duerr09,norsen14b,bri,tumulka22}, the spin is co-determined by the system and the details of the experimental setup.

In Section \ref{sec:equations}, we introduce the Bohmian evolution equations for an electron in an electromagnetic field.
We then apply these equations to a Stern-Gerlach type setup in Section \ref{sec:sterngerlach}.
We consider an entangled 2-electron system in Section \ref{sec:sg_entangled} and compare the situation where both particles are evolving freely with the situation where one of the electrons is sent through a Stern-Gerlach device.
There we visualize how this affects the zig-zagging of the other electron as an action-at-a-distance.

\section{Zig-zag dynamics}\label{sec:equations}
In the nonrelativistic limit \cite{struyve12}, the zig-zag dynamics for a single electron describes the change of  its position ${\bf X}$ and chirality $\chi=\pm 1$ by 
\begin{equation} \label{eq:langevin}
\dot {\bf X}(t) = {\bf v}_{\chi(t)}\big({\bf X}(t),t\big), \qquad \chi (t)\rightarrow -\chi(t) \;\text{ at rate } \, r_{-\chi(t)}\big({\bf X}(t),t\big),
\end{equation}
where the velocity ${\bf v}_{\chi}({\bf x},t)$ and jump rate $r_{\chi}({\bf x},t)$ are determined by the Pauli spinor $\Psi({\bf x},t)$ as follows,
\begin{align}
 {\bf v}_\chi & = \frac{\hbar}{m} \frac{\Im{(\Psi^\dagger {\bf D} \Psi)}}{\Psi^\dagger \Psi} + \frac{\hbar}{2m} \frac{{\boldsymbol{ \nabla}}\times\left(\Psi^\dagger\boldsymbol{\sigma}\Psi\right)}{\Psi^\dagger\Psi} + c\, \chi\, {\bf s} \label{eq:vc}\\[.8em]
 r_{\chi} & = \left[ c\,\chi\, \frac{{\boldsymbol{\nabla}}\cdot\left(\Psi^\dagger\boldsymbol{\sigma}\Psi\right)}{\Psi^\dagger\Psi} \right]^+  = \left[ 2c \chi \frac{\Re{\Psi^\dagger \boldsymbol{\sigma}\cdot{\boldsymbol \nabla}\Psi}}{\Psi^\dagger \Psi}\right]^+,\quad F^+  ={\textrm{max}}(F,0) ,\label{eq:tum}
\end{align}
where ${\bf D}  = {\boldsymbol{\nabla}} - \ii e {\bf A}/\hbar c$ is the covariant derivative with vector potential $\bf A$, and 
\begin{equation}
{\bf s}  = \frac{\Psi^\dagger\boldsymbol{\sigma}\Psi}{\Psi^\dagger \Psi},
\label{spin vector}
\end{equation}
with $\boldsymbol{\sigma}$ the vector of the three Pauli matrices, is the spin vector.  
The Pauli spinor satisfies the Pauli equation, which in an external electromagnetic potential $(V,{\bf A})$ with corresponding magnetic field~${\bf B}$, reads 
\begin{equation}\label{eq:pauli}
\ii \hbar \pa_t \Psi = -\frac{\hbar^2}{2m} {\bf D}^2 \Psi - \frac{e\hbar}{2mc} {\bf B}\cdot {\boldsymbol{\sigma}} \Psi + e V\Psi .
\end{equation}
The Fokker-Planck equation associated to the stochastic dynamics \eqref{eq:langevin} is
\be\label{eq:rtp}
\pa_t \rho({\bf x},\chi,t) + {\boldsymbol \nabla} \cdot ({\bf v}_\chi  \rho({\bf x},\chi,t)) = r_{\chi} \rho({\bf x},-\chi,t) - r_{-\chi} \rho({\bf x},\chi,t).
\en
A particular solution is given by $\rho({\bf x},\chi,t) = \Psi^\dagger \Psi/2$, since the Pauli equation \eqref{eq:pauli} implies  
\begin{equation}\label{eq:rtp2}
\pa_t (\Psi^\dagger \Psi) + {\boldsymbol \nabla} \cdot ({\bf v}_\chi\Psi^\dagger \Psi) = r_{\chi} \Psi^\dagger \Psi - r_{-\chi} \Psi^\dagger \Psi .
\end{equation}
Therefore, the dynamics \eqref{eq:langevin} preserves the \emph{quantum equilibrium distribution} $\Psi^\dagger \Psi$, which is important in establishing the usual quantum predictions determined by the Born rule \cite{duerr92a,duerr09}.
(In the relativistic case, the equilibrium distribution $\rho({\bf x},\pm 1,t)$ is determined by the right- and left-handed Weyl spinor $\psi_{\pm}$ and hence the right- and left-handed part are generically different.
In the nonrelativistic limit, the chiral components become equal, at least to the lowest order.)

Note that \eqref{eq:langevin} is a Langevin-type equation, and may be recognized as describing run-and-tumble dynamics, with the spacetime profile of the Pauli spinor playing the role chemotaxis does for run-and-tumble bacteria \cite{maes22}.
The Fokker-Planck equation for the density of two-state run-and-tumble particles is precisely \eqref{eq:rtp}.

The chirality jumps generate the zig-zag motion of the particles.
The chirality dependence of the velocity ${\bf v}_\chi$ is only in the third term of \eqref{eq:vc}, which is proportional to the spin vector.
This term is also the dominating contribution to the velocity since its norm is given by the light speed $c$.
As a consequence, the particle travels approximately at the speed of light and approximately in the direction of the spin vector. (In the relativistic case, the velocity is \emph{exactly} along the direction of the spin vector corresponding to the left- or right-handed Weyl spinor and the motion is always luminal \cite{struyve12}.)

In earlier formulations of a Bohmian dynamics for the Pauli theory, there was no stochasticity.
In the simplest formulation, the velocity of the particle is given by only the first term of \eqref{eq:vc} \cite{bohm93,holland93b,duerr09}.
The second term was motivated by considering the nonrelativistic limit of the Bohmian dynamics for the Dirac theory \cite{bohm93}.
There are already striking differences resulting from the addition of this second term, for example in hydrogenic atoms \cite{colijn02} and in the double slit experiment \cite{holland03a}.
The inclusion of the third term and the jumps entails further significant changes, as illustrated in \cite{maes22} for the double-slit experiment.
One immediate consequence is that the particle can never be at rest, unlike those earlier formulations where the particle remains at rest for certain energy eigenstates (which was a feature that bothered Einstein \cite{einstein53,myrvold03}).

For a multiple-electron system, the Pauli equation for $\Psi({\bf x}_1,\dots,{\bf x}_N,t)$ is
\begin{equation}
\ii \hbar \pa_t \Psi =
 \sum^N_{k=1}\left[- \frac{\hbar^2}{2m} {\bf D}_k^2 \Psi -  \frac{e\hbar}{2m c} {\bf B}({\bf x}_k)\cdot \boldsymbol{\sigma}_k \Psi +  e  V({\bf x}_k) \Psi\right],
\label{eq:pauli_many}
\end{equation}
where now  ${\bf D}_k =  \boldsymbol{\nabla}_k - \ii e {\bf A} ({\bf x}_k)/\hbar c$.
The corresponding zig-zag dynamics for the $k$th particle is 
\begin{equation}
\dot {\bf X}_k(t) = {\bf v}_{k,\chi_{k}(t)}\big({\bf X}_1(t),\dots,{\bf X}_N(t),t\big),
\end{equation}
with
\begin{equation}
{\bf v}_{k,\chi_k} = \frac{\hbar}{m} \frac{\Im{(\Psi^\dagger {\bf D}_k \Psi)}}{\Psi^\dagger \Psi}  + \frac{\hbar}{2m} \frac{\boldsymbol{\nabla}_k\times\left(\Psi^\dagger\boldsymbol{\sigma}_k\Psi\right)}{\Psi^\dagger\Psi} + c \chi_k {\bf s}_k ,
\label{eq:v_many}
\end{equation}
where ${\boldsymbol \sigma}_k  = I \otimes \cdots \otimes I \otimes {\boldsymbol \sigma} \otimes I \otimes \cdots \otimes I$, with ${\boldsymbol \sigma}$ at the $k$-th of the $N$ places, and 
\be
{\bf s}_k = \frac{\Psi^\dagger\boldsymbol{\sigma}_k\Psi}{\Psi^\dagger\Psi}
\en
is the spin vector for the $k$th particle. The jump rate for one of the chiralities, say the $k$th, to flip from $-\chi_k$ to $\chi_k$ at time $t$ is given by 
\begin{equation}
r_{k,\chi_k} \big({\bf X}_1(t),\dots,{\bf X}_N(t),t\big),
\end{equation}
with
\begin{equation}
r_{k,\chi_k} = \left[ c \chi_k \frac{{\boldsymbol{\nabla}}_k\cdot\left(\Psi^\dagger\boldsymbol{\sigma}_k\Psi\right)}{\Psi^\dagger\Psi} \right]^+.
\label{eq:rates_many}
\end{equation}
It is crucial here that ${\bf v}_{k,\chi_k}$  and $r_{k,\chi_k}$ depend on the full wave function (multi-particle spinor $\Psi$), which implies that the dynamics of one particle may nonlocally depend on the motion of the other particles.
In particular, the spin vector and hence the (approximate) direction of zig-zagging may depend on the whole configuration.

Unlike the single-particle case, the velocity need not be luminal (which is also true in the full relativistic theory).
This is because for an entangled spinor, the norm of the spin vector ${\bf s}_k$ may be smaller than one.
We will encounter examples in Section \ref{sec:sg_entangled}.

\section{Stern-Gerlach setup}\label{sec:sterngerlach}
\subsection{Solution of the Pauli equation}

In the Stern-Gerlach setup, a beam of separate spinful particles is sent through an inhomogeneous magnetic field, resulting in a splitting of the beam and two detection bands on a final detecting screen. Nonrelativistically, such particles are described by the Pauli equation \eqref{eq:pauli}, which in natural units $\hbar = m = c = 1$, is given by 
\begin{equation}
\ii \pa_t \Psi = - \frac{1}{2}  \nabla^2 \Psi -  \frac{e}{2}{\bf B}\cdot {\boldsymbol{\sigma}} \Psi  .
\label{eq:weak_pauli}
\end{equation}
Rather than taking the magnetic field to be spatially varying in the longitudinal direction, we will assume for simplicity a time-dependent variation, so that the magnetic field is nonzero only between some initial and final time $t_i$ and $t_f$, approximately corresponding to the time span the bulk of the packet passes through the region of the magnets, {\it i.e.},
\begin{equation}
    \mathbf{B}(t)  = \left\{\begin{array}{lll}
        (0,\ 0,\ 0) && {\mathrm{if}} \ t<t_i, \\
        (0,\ 0,\ 2 b z/e) &&  {\mathrm{if}} \ t_i<t<t_f,\\
        (0,\ 0,\ 0) && {\mathrm{if}} \ t>t_f.
    \end{array}\right. 
\end{equation}
This will suffice for our qualitative modeling of the spin polarization. 


\begin{equation} \textbf{B}( t) = \left\{\begin{array}{lll} 
( 0, \ 0, \ 0) && {\mathrm{if}}\ \ t<t_i, \\
        (0,\ 0,\ 2 b z/e) &&  {\mathrm{if}} \ t_i<t<t_f,\\
        (0,\ 0,\ 0) && {\mathrm{if}} \ t>t_f. \end{array}\right. \tag{3.2}\end{equation}

We have also adopted the usual approximations where the vector potential is dropped in the kinetic term and the magnetic field does not satisfy Maxwell's equations, since ${\boldsymbol \nabla} \cdot {\bf B} \neq 0$. While this can be motivated in the case of atom beams, see {\it e.g.}\ \cite[pp.\ 328-330]{boehm86}, it is not justified for electron beams. Namely, because of the charge of the electron, the vector potential cannot be dropped in the kinetic term. It is responsible for a Lorentz force which even blurs the intended spin polarization \cite{mott29,pauli32}. However, other possibilities have been suggested for implementing the polarization of electron beams \cite{batelaan97,garraway99,mcgregor15}. We will not explore any of these possibilities here, but merely adopt the above dynamics as a toy model for the polarization. The advantage is that this modeling allows us to use analytic solutions \cite{hsu11}. Actual implementations of the polarization should yield qualitatively similar features.

Consider the initial Pauli spinor
\begin{equation}\label{eq:initial_wf_cs}
\Psi({\bf x},0) =  \psi({\bf x},0)\begin{pmatrix} 
c_+ \\
c_- 
\end{pmatrix},
\end{equation}
with $c_\pm \in \mathbb{C}$, and where the scalar wave function 
\begin{equation}
\psi({\bf x},0) = \psi_x(x,0)\psi_y(y,0)\psi_{z}(z,0)
\end{equation}
is Gaussian with momentum $p$ in the $x$-direction, and has widths $d_x,d_y,d_z$ in the various directions, {\it i.e.},
\begin{equation}
    \begin{gathered}
        \psi_x(x,0) = \frac{1}{\left(2\pi d_x^2 \right)^{1/4}} \exp\left(-\frac{x^2}{4d_x^2 }+\ii p x \right),\\[1em]
        \psi_y(y,0) = \frac{1}{\left(2\pi d_y^2 \right)^{1/4}} \exp\left(-\frac{y^2}{4d_y^2 }\right) ,\qquad \psi_z(z,0) = \frac{1}{\left(2\pi d_z^2 \right)^{1/4}} \exp\left(-\frac{z^2}{4d_z^2 }\right) .
    \end{gathered}
\end{equation}
Using the propagator found in \cite{hsu11}, the solution of \eqref{eq:weak_pauli} is 
\begin{equation}
\Psi({\bf x},t) = \begin{pmatrix} 
c_+\psi_+({\bf x},t) \\
c_-\psi_- ({\bf x},t)
\end{pmatrix},
\label{eq:solution_pauli_1}
\end{equation}
where 
\begin{equation}
    \psi_\pm({\bf x},t)= \psi_x(x,t)\psi_y(y,t)\psi_{z,\pm}(z,t),
    \label{eq:solution_pauli_2}
\end{equation}
with
\begingroup
\addtolength{\jot}{0.em}
\begin{align}\nonumber
    \psi_x(x,t) &= \frac{1}{\left[2\pi d_x^2 \left(1 + \frac{\ii t}{2d_x^2}\right)^2\right]^{1/4}} \exp\left[-\frac{(x-p t)^2}{4d_x^2 \left(1 + \frac{\ii t}{2d_x^2}\right)}+\ii p x - \ii \frac{p^2}{2} t\right],\label{eq:solution_pauli_3}\\
    \psi_y(y,t) &= \frac{1}{\left[2\pi d_y^2 \left(1 + \frac{\ii t}{2d_y^2}\right)^2\right]^{1/4}} \exp\left[-\frac{y^2}{4d_y^2 \left(1 + \frac{\ii t}{2d_y^2}\right)}\right],\\\nonumber
    \psi_{z,\pm}(z,t) &= \frac{1}{\left[2\pi d_z^2 \left(1 + \frac{\ii t}{2d_z^2}\right)^2\right]^{1/4}}\\
    &\times\left\{
        \begin{aligned}
            &\exp\left[-\frac{z^2}{4d_z^2 \left(1 + \frac{\ii t}{2d_z^2}\right)}\right] &{\mathrm{if}} \ t < t_i\\
            &\exp\left[-\frac{\left(z \mp \frac{b (t-t_i)^2}{2}\right)^2}{4d_z^2 \left(1 + \frac{\ii t}{2d_z^2}\right)} \pm \ii b (t-t_i) z - \ii \frac{b^2 (t-t_i)^3}{6}\right] &{\mathrm{if}} \ t_i < t < t_f\\
            &\begin{aligned}
                \,\exp&\left[-\frac{\left(z \mp \frac{b (t_f-t_i)^2}{2} \pm b(t_f-t_i)(t-t_f)\right)^2}{4d_z^2 \left(1 + \frac{\ii t}{2d_z^2}\right)} \right. \\
                &\hspace{1em}\pm \ii b (t_f-t_i) z - \ii \frac{b^2 (t_f-t_i)^3}{6}- \ii \frac{b^2 (t_f-t_i)^2}{2}(t-t_f)\Bigg] 
            \end{aligned} &{\mathrm{if}} \ t_f < t 
        \end{aligned}\right.  .\nonumber
\end{align}
\endgroup

\subsection{Integrating the zig-zag dynamics}

To numerically integrate the trajectories corresponding to the wave function \eqref{eq:solution_pauli_3}, we make use of the code developed for our previous study \cite{maes22} (extending it to allow for wave functions that are not spin eigenstates).

The propagation of the particle consists of two parts. First, the deterministic part of the dynamics \eqref{eq:langevin} is given by a first-order ordinary differential equation, which we integrate using the Cash-Karp method, a Runge-Kutta method which yields a fifth-order accurate solution, as well as a fourth-order estimate for the error. This estimate is then used to adaptively adjust the time step to keep the local truncation error low, for which we set an absolute tolerance of $10^{-10}$.
At this point, we also make sure that the product of the tumbling rate and the time step remains reasonably low, below $2^{-7}$.
This product is then taken to be the probability for the particle to undergo a chirality jump during this time step.
This integration step is repeated until the final time is reached.
See \cite{zigzag2} for the publicly available code.

The parameters (in natural units) chosen to integrate the zig-zag dynamics  for the spinor~\eqref{eq:solution_pauli_1} are $d_x = d_y = d_z = 100$, $b = 10^{-6}$, and $p = 1/10$. While these parameters are not experimentally feasible (for example, in standard units $d_x = d_y = d_z \approx \SI{40}{\pico\meter}$), they allow for a clear visualization which captures the qualitative features of the dynamics.

\begin{figure}
    \centering
    \begin{subfigure}{.9\textwidth}
        \centering
        \includegraphics[width=\textwidth]{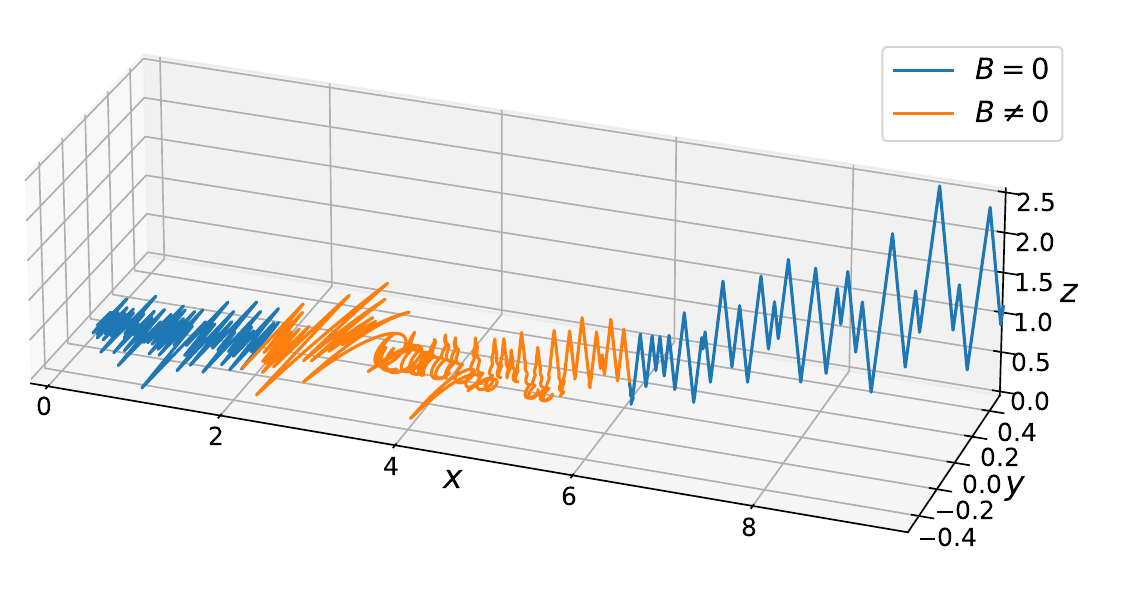}
        \caption{}
        \label{fig:sg_spiny_3d}
    \end{subfigure}
    \begin{subfigure}{.9\textwidth}
        \centering
        \includegraphics[width=\textwidth]{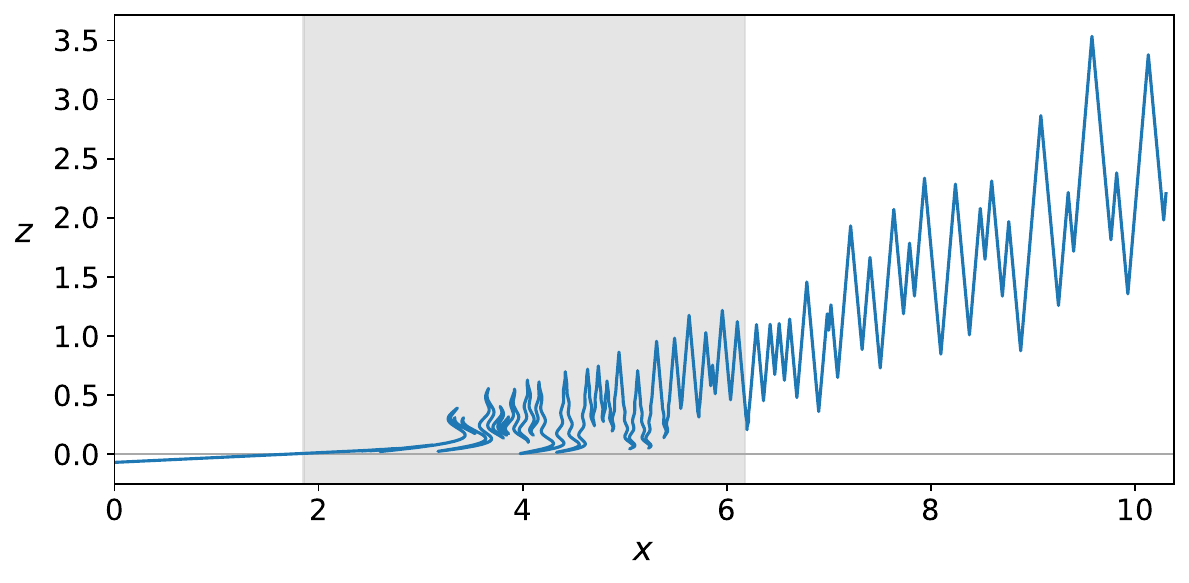}
        \caption{}
        \label{fig:sg_spiny}
    \end{subfigure}
     \caption{Particle trajectory for the spinor $\Psi = \frac{1}{\sqrt{2}}(\psi_+,  \ii \psi_-)^\tran$, which has initial spin {\em up} in the $y$-direction. Fig.~\ref{fig:sg_spiny} shows the $xz$-projection of Fig.~\ref{fig:sg_spiny_3d}. The area where the magnetic field is turned on is marked by a different line colour and a grey background in Figs.~\ref{fig:sg_spiny_3d}~\&~\ref{fig:sg_spiny} respectively. As is clear from the figure, the particle may cross the $xy$-plane.}
     \label{fig:spiny}
\end{figure}

The total running time is $T = 10^5$ and the magnetic field is turned on between $t_i = T/5$ and  $t_f=3T/5$. 
The space and time coordinates in all the figures have been rescaled with a factor of $10^3$ to reduce visual clutter on the axes.

\begin{figure}
    \begin{subfigure}{.495\textwidth}
        \centering
        \includegraphics[width=\textwidth]{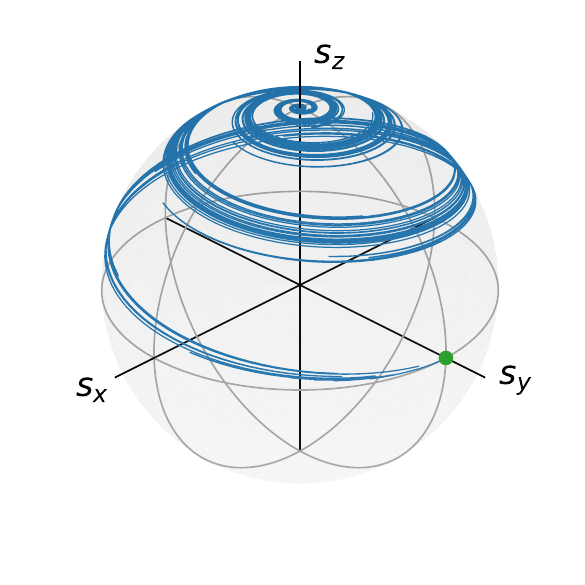}
        \caption{}
        \label{fig:spin3d}
    \end{subfigure}
    \hfill
    \begin{subfigure}{.45\textwidth}
        \centering
        \includegraphics[width=\textwidth]{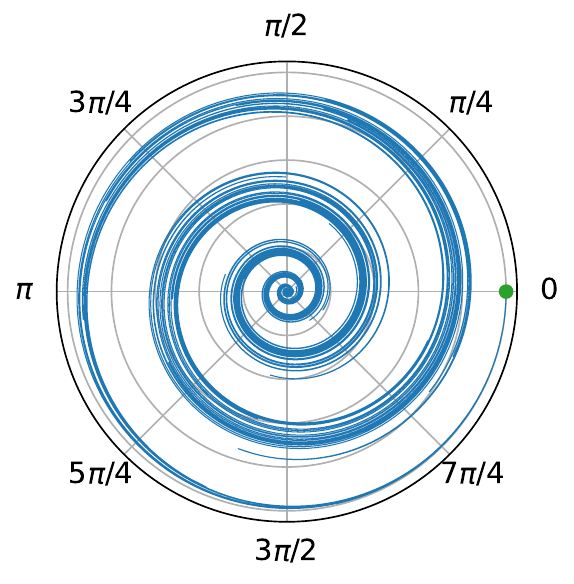}
        \caption{}
        \label{fig:spin2d}
   \end{subfigure}
   \caption{Spin vector along the trajectory plotted in Fig.~\ref{fig:spiny}.
   Fig.~\ref{fig:spin3d} shows the evolution of the spin vector on the unit sphere, which can be identified with the Bloch sphere, while Fig.~\ref{fig:spin2d} shows its projection onto the $xy$-plane. It shows the transitioning of the spin vector from the positive  $y$-direction to the positive $z$-direction. The winding up and down the sphere is a direct consequence of the particle zig-zagging back and forth in position space, where the spin vector varies spatially, cf. Fig.~\ref{fig:spinvector}.}
   \label{fig:spinyspin}
\end{figure}

In Fig.~\ref{fig:spiny}, a trajectory is plotted for a spinor that initially has spin {\it up} in the $y$-direction, {\it i.e.}, $\Psi = \frac{1}{\sqrt{2}}(\psi_+,  \ii \psi_-)^\tran$, so that $c_+ = 1/\sqrt{2}$ and $c_- = \ii/\sqrt{2}$. 
The spin vector ${\bf s}$ along this trajectory is plotted in Fig.~\ref{fig:spinyspin}. 
As explained before, the zig-zagging happens approximately along the direction of the spin vector.
From Fig.~\ref{fig:spiny} it is clear that the zig-zagging which is initially along the $y$-direction transitions into a zig-zagging along the $z$-direction as it passes the magnetic field. As shown in Fig.~\ref{fig:spinyspin}, this transition involves some heavy oscillation (due to the jumps) before settling in the $z$-direction. This happens quite fast, with changes occurring only while the particle moves through the magnetic field.

\begin{figure}
    \begin{subfigure}{.49\textwidth}
        \centering
        \includegraphics[width=\textwidth]{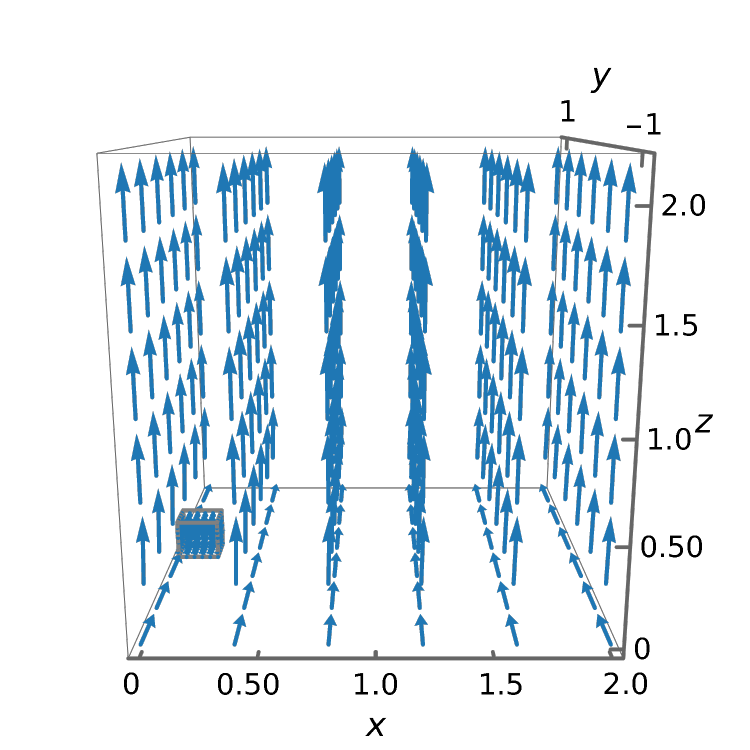}
        \caption{}
        \label{fig:pijltjes1}
    \end{subfigure}
    \hfill
    \begin{subfigure}{.49\textwidth}
        \centering
        \includegraphics[width=\textwidth]{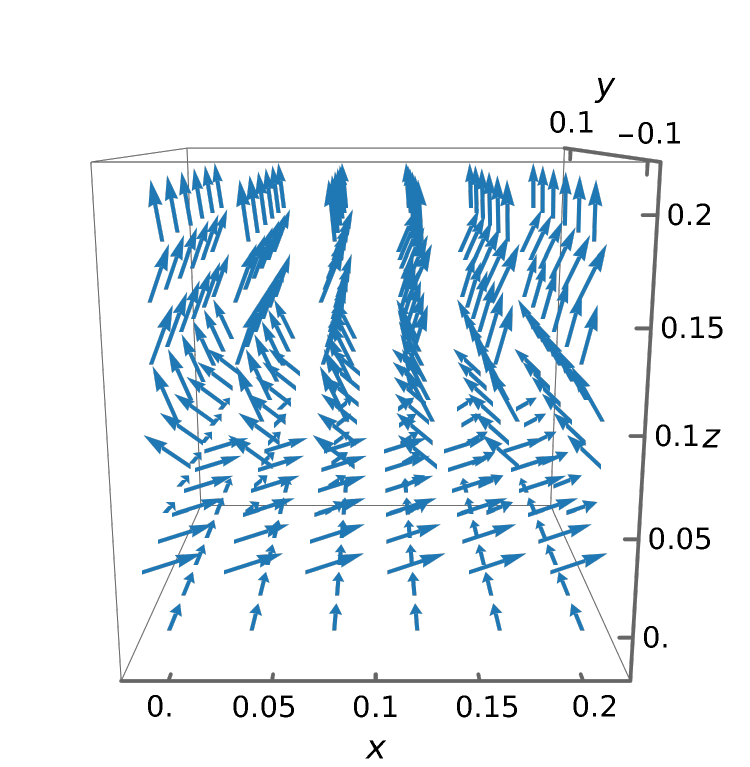}
        \caption{}
        \label{fig:pijltjes2}
   \end{subfigure}
   \caption{The spin vector above the $xy$ plane ($z\geqslant0$), at $t=7\cdot 10^4$, just after turning off the magnetic field, for the spinor $\Psi = \frac{1}{\sqrt{2}}(\psi_+,  \ii \psi_-)^\tran$, as in Figs.~\ref{fig:spiny} and \ref{fig:spinyspin}. Fig.~\ref{fig:pijltjes2} zooms in on part of Fig.~\ref{fig:pijltjes1} shown in the bottom left corner.  Away from the $xy$-plane, the spin vector points in the positive $z$-direction, but near the plane it displays irregular behaviour, cf. Fig~\ref{fig:spinyspin}.}
   \label{fig:spinvector}
\end{figure}

The spin vector at time $t=7\cdot 10^4$ (just after turning off the magnetic field) is plotted in Fig.~\ref{fig:spinvector}. 
In the region sufficiently above (below) the $xy$-plane the spin vector points in the positive (negative) $z$-direction. 
Near the $xy$-plane the spin vector shows more irregular behaviour, which explains the strong oscillation of the spin vector along the trajectory shown in Fig.~\ref{fig:spinyspin}. 
On the $xy$-plane, the spin vector points in the positive $y$-direction. 
So once the particle comes near this plane it will approximately start moving and zig-zagging along the $y$-direction.

\begin{figure}
    \centering
    \includegraphics[width=\textwidth]{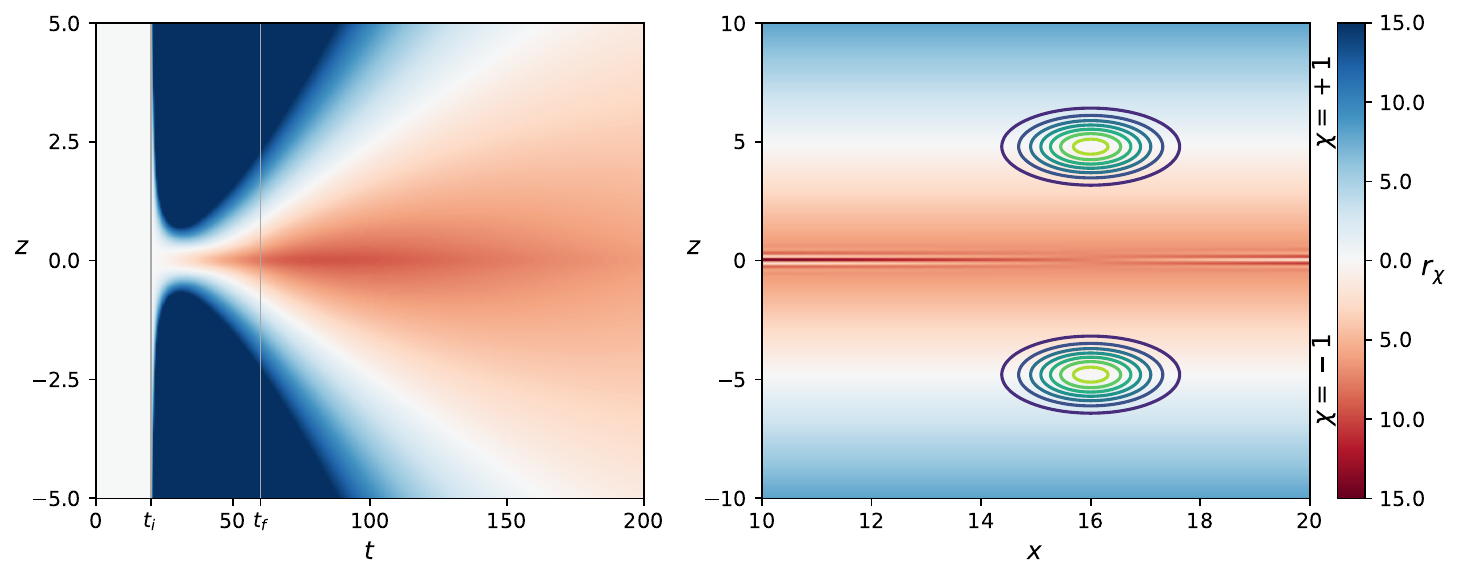}
    \caption{
    Left: Time-dependent tumbling rates for the spinor $\Psi = \frac{1}{\sqrt{2}}(\psi_+,  \ii \psi_-)^\tran$ as a function of $z$ at $y=0$ with the $x$-coordinate at the centre of the wave packet.
    Right: Tumbling rates at time $160$, at $y=0$. The contour lines connect equal densities $\Psi^\dagger\Psi(x,0,z,160)$.
    }
    \label{fig:tumblerate}
\end{figure}

Whenever the particle moves away from the bulk of the wave packet, the jump rate increases, increasing the probability that it changes chirality, and thus direction, and moves back towards the center of the packet. This is clear from Fig.~\ref{fig:tumblerate}, where the jump rates are plotted. 
So when a particle is above the $xy$-plane and moving down towards it, the probability for a jump increases, and with it, the probability to move back up. If no jump occurs and the particle gets near the $xy$-plane it will start moving approximately in the $y$-direction, where again the probability for a jump will increase when the particle moves further away from the bulk of the packet. 
Once the jump occurs, it will reverse its motion and will start moving up again eventually. 
While a particle may cross the $xy$-plane the probability for this to happen decreases over time, because of the separation of the packets $\psi_+$ and $\psi_-$.

Depending on the initial position and the jumps that took place, the final position ends up either above or below the $xy$-plane, which determines the outcome of the spin measurement by the Stern-Gerlach device.

The trajectory is markedly different from those in Fig.~\ref{fig:sg_standard}, which follow the original Bohmian dynamics for the same spinor. 

\begin{figure}
    \centering
    \includegraphics[width=0.9\textwidth]{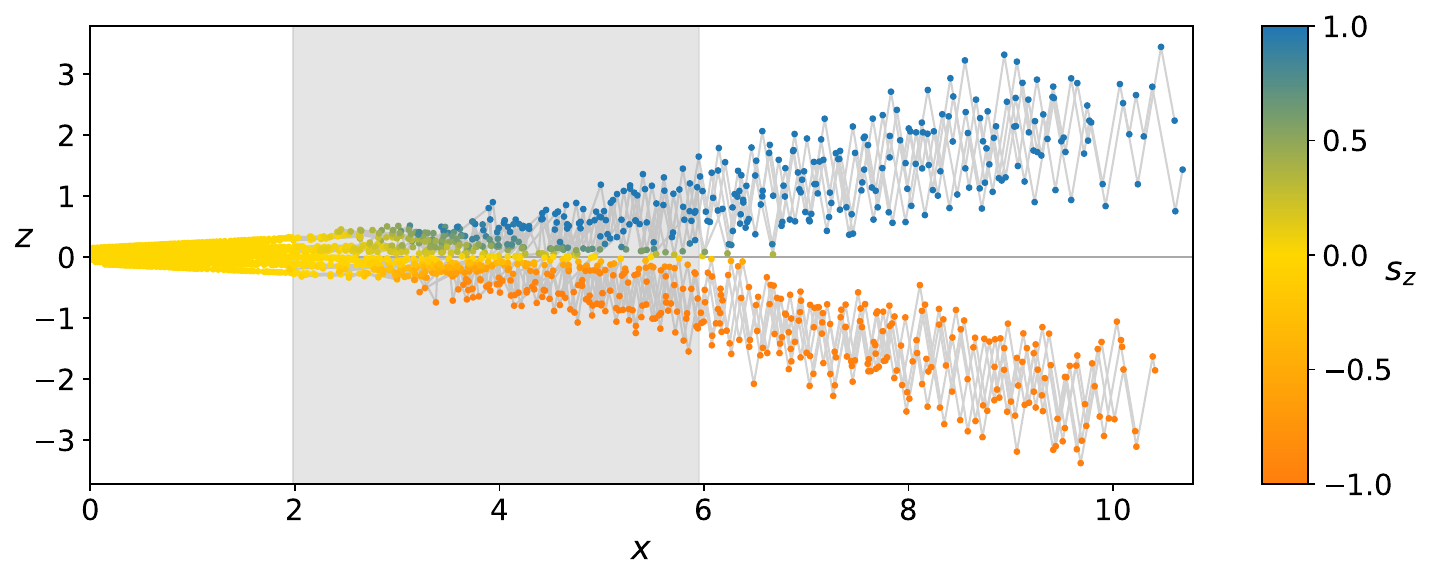}
    \caption{Projection in the $xz$-plane of 10 particle trajectories for the spinor $\Psi = \frac{1}{\sqrt{2}}(\psi_+,  \ii \psi_-)^\tran$. The dots represent the locations of the particles at the times of jumps. The color of a dot indicates $s_z$, the $z$-component of the spin vector, at the jump event.}
    \label{fig:pastelsg}
\end{figure}

In Fig.~\ref{fig:pastelsg}, 10 trajectories are plotted, with initial positions randomly drawn with a $\Psi^\dagger \Psi$ distribution, again for the state $\Psi = \frac{1}{\sqrt{2}}(\psi_+,  \ii \psi_-)^\tran$, which has initial spin along the y-direction.
The locations where jumps occur are represented by a dot.
The color of the dot represents the value of $s_z$, the $z$-component of the spin vector, at those locations, at the time of the chirality flip.

\begin{figure}
    \centering
    \includegraphics[width=0.9\textwidth]{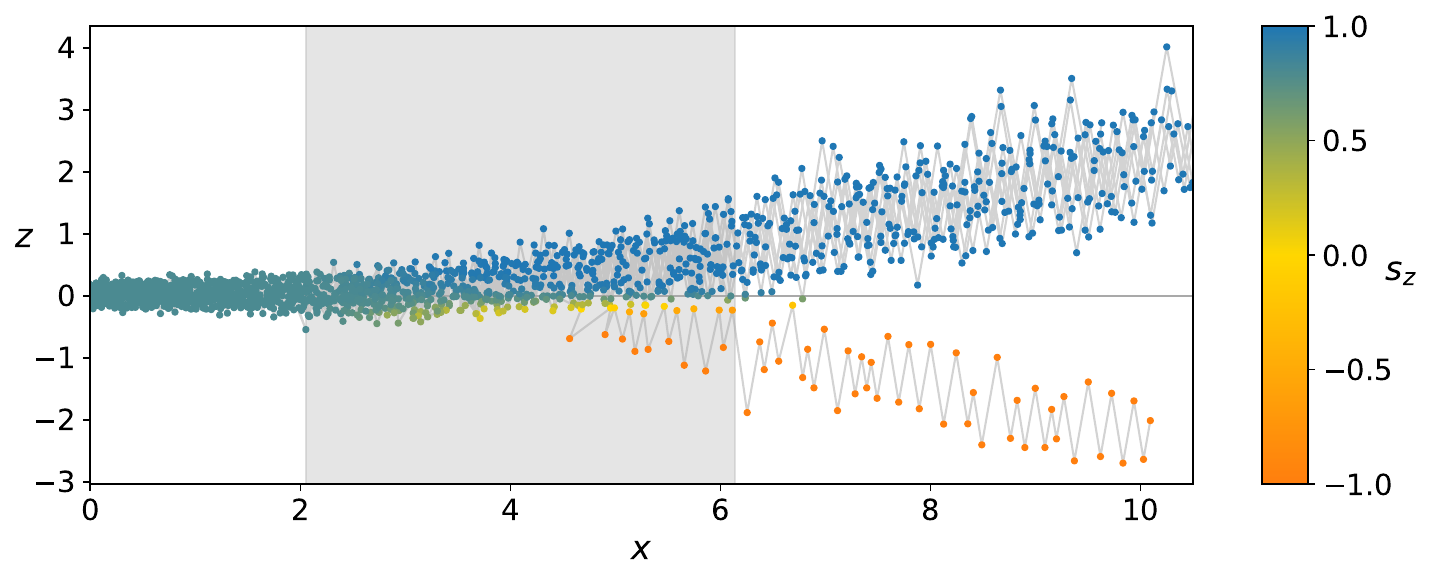}
    \caption{Projection in the $xz$-plane of 10 particle trajectories for the spinor $ \Psi = \frac{1}{\sqrt{10}}\left(3 \psi_+,\ii \psi_-\right)^\tran $. As in Fig.~\ref{fig:pastelsg}, the dots represent the locations of the particles at the times of jumps. The color of a dot indicates $s_z$ at the jump event.}
    \label{fig:sg}
\end{figure}

In Fig.~\ref{fig:sg}, similarly, trajectories are plotted for the wave function $ \Psi = \frac{1}{\sqrt{10}}\left(3 \psi_+,\ii \psi_-\right)^\tran $.
In this case, the initial spin vector is $(0,6/10,8/10)$. More trajectories end up in the upper beam. This reflects the usual quantum probability of 9/10 for the particle to be found with spin {\em up} in the $z$-direction. The chirality jumps play a role in deciding whether the particle ends up with a positive or negative $z$-coordinate.

As mentioned before, the separation of the waves $\psi_+$ and $\psi_-$, which happens after passage through the magnets, implies that the particle tends to stay within the bulk of either the wave $\psi_+$ or $\psi_-$ and does not tend to move between $\psi_+$ and $\psi_-$.\footnote{The fact that the particle does not tend to move between supports of non-overlapping wave functions, will even be amplified for macroscopic objects. So, in particular, this will imply stability of macroscopic records, which is important in showing that the theory is empirically adequate \cite{duerr92a}.}
This also implies an {\it effective collapse}. Namely, if the particle goes in the positive $z$-direction after passing the magnets, its dynamics is effectively determined only by $\psi_+$ (since the velocity and the jump rate only depend on the wave function in the neighborhood of the particle).
This means that (as long as the supports of $\psi_+$ and $\psi_-$ remain disjoint, and the particle does not cross the $xy$-plane), the $\psi_-$ part can be ignored in the description of the particle, amounting to an effective collapse of the spinor to $(\psi_+,0)^\tran$.\footnote{There is actually a more precise notion of collapse in Bohmian mechanics. This requires the notion of the wave function of a subsystem, which can be naturally defined in terms of the wave function of the universe and the particle positions of the environment \cite{duerr92a}. In the present case, which deals with spin, the state of a subsystem is a density matrix, called the {\em conditional density matrix}, rather than a wave function, and this density matrix will actually collapse in measurement situations \cite{duerr05b}.}

In summary, the outcome of the spin measurement depends on where the particle ends up after passing through the Stern-Gerlach device. Where it ends up depends not only on the initial position but also on the jumps that have occurred. Furthermore, if at $t=0$ the particle position has distribution $\Psi^\dagger ({\bf x},0)\Psi({\bf x},0)$, then at time $t$ it has distribution $\Psi^\dagger({\bf x},t) \Psi({\bf x},t)$, in accordance with the Born rule.

\section{Stern-Gerlach with a pair of entangled particles}\label{sec:sg_entangled}

When there are no electromagnetic potentials, a particular solution of the 2-particle Pauli equation \eqref{eq:pauli_many} is given by the entangled spinor
\begin{equation}
\Psi({\bf x}_1,{\bf x}_2,t) = \psi_1({\bf x}_1,t) \psi_2({\bf x}_2,t) \left[ a \begin{pmatrix}  1 \\ 0 \end{pmatrix}\begin{pmatrix}  0 \\ 1 \end{pmatrix}  - b \begin{pmatrix}  0 \\ 1 \end{pmatrix} \begin{pmatrix}  1 \\ 0 \end{pmatrix}\right],
\end{equation}
where $a$ and $b$ are real constants, and where $\psi_1$ and $\psi_2$ are scalar wave functions for particles 1 and 2 respectively, satisfying the free Schr\"odinger equation.
We have that
\begin{equation}
\Psi^\dagger\boldsymbol{\sigma}_1\Psi = |\psi_1|^2  |\psi_2|^2 (a^2 - b^2) {\bf e}_z = -\Psi^\dagger\boldsymbol{\sigma}_2\Psi.
\end{equation}
So the spin vectors of the two particles will always be pointing in opposite directions.
In addition, the velocity fields \eqref{eq:v_many} for the two particles reduce to
\begin{equation}
{\bf v}_{k,\chi_k} = \frac{\hbar}{m} \frac{\Im{\left(\psi^*_k {\boldsymbol{\nabla}}_k \psi_k\right)}}{|\psi_k|^2}  - (-1)^k \frac{a^2 -b^2 }{a^2 + b^2} \left[\frac{\hbar}{2m} {\boldsymbol{\nabla}}_k\times\left(\ln|\psi_k|^2 {\bf e}_z\right) + c \chi_k\right] , \qquad k=1,2,
\end{equation}
and the jump rates \eqref{eq:rates_many} become
\begin{equation}
r_{\chi_k} = \left[- c\chi_k (-1)^k \frac{a^2 -b^2 }{a^2 + b^2} \pa_{z_k}\ln|\psi_k|^2 \right]^+, \qquad k=1,2.
\end{equation}
As such, the spin-dependent part in both the velocity and the jump rates decreases when $a$ approaches $b$. 
In the special case that $a=b$, the spin-dependent part completely vanishes,\footnote{For a single particle, this is impossible since its spin vector \eqref{spin vector} has norm one and never vanishes.} so that the dynamics becomes
\begin{equation}
\dot {\bf X}_k = \frac{\hbar}{m} \frac{\Im{(\psi^*_k {\boldsymbol{\nabla}}_k \psi_k)}}{|\psi_k|^2}. 
\label{original}
\end{equation}
That is, the dynamics amounts to the original one introduced by de\ Broglie and Bohm, and the particles also move entirely independently. 

As a particular example of the case where $a=b$, we can consider $\psi_1$ and $\psi_2$ to be Gaussians initially localized at the origin which are moving away from each other.
That is, we take particle~1 with positive momentum $p$ and particle~2 with negative momentum $-p$ along the $x$-axis. Denoting the solution \eqref{eq:solution_pauli_1} 
by 
\begin{equation}
\Psi_{p,{\bf B}}({\bf x},t) = \frac{1}{{\sqrt 2}}\begin{pmatrix} 
\psi_{p,{\bf B},+}({\bf x},t) \\
\psi_{p,{\bf B},-} ({\bf x},t)
\end{pmatrix},
\label{sol5}
\end{equation}
the spinor corresponding to the two particles moving freely apart (${\bf B}={\bf 0}$) is 
\begin{equation}
\Psi_{\textrm{free}}({\bf x}_1,{\bf x}_2,t) = 
\frac{1}{{\sqrt 2}}\psi_{p,{\bf 0},+}({\bf x}_1,t) \psi_{-p,{\bf 0},+}({\bf x}_2,t) \left[  \begin{pmatrix}  1 \\ 0 \end{pmatrix}\begin{pmatrix}  0 \\ 1 \end{pmatrix}  -  \begin{pmatrix}  0 \\ 1 \end{pmatrix} \begin{pmatrix}  1 \\ 0 \end{pmatrix}\right],
\end{equation}
where 
$\psi_{p,{\bf 0},+} = \psi_{p,{\bf 0},-}$.
For this state, the particle dynamics is of the form \eqref{original}, without the zig-zagging. 

\begin{figure}[H]
    \centering
    \includegraphics[width=0.75\textwidth]{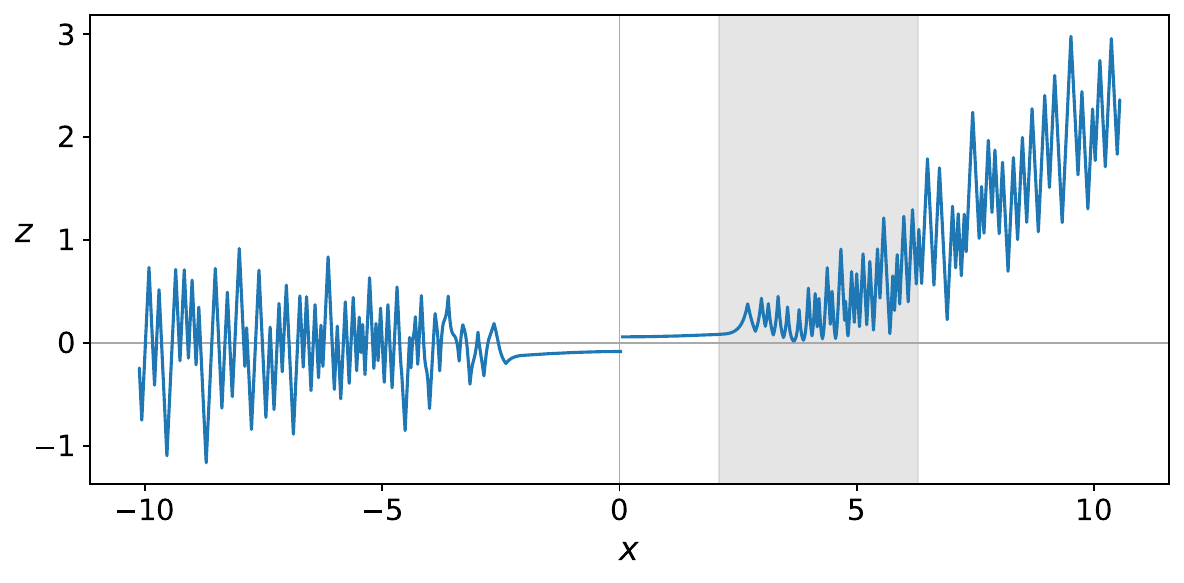}
    \caption{A set of trajectories for the entangled spinor $\Psi_{\textrm{SG}}$ given in \eqref{eq:entangled}, projected in the $xz$-plane. Initially, the particles are moving without zig-zagging. The zig-zagging of both particles starts once particle~1 enters the Stern-Gerlach device. After passing through the device the zig-zagging is along the $z$-direction for both particles.}
    \label{fig:sg1}
\end{figure}

\begin{figure}[H]
    \centering
    \includegraphics[width=0.85\textwidth]{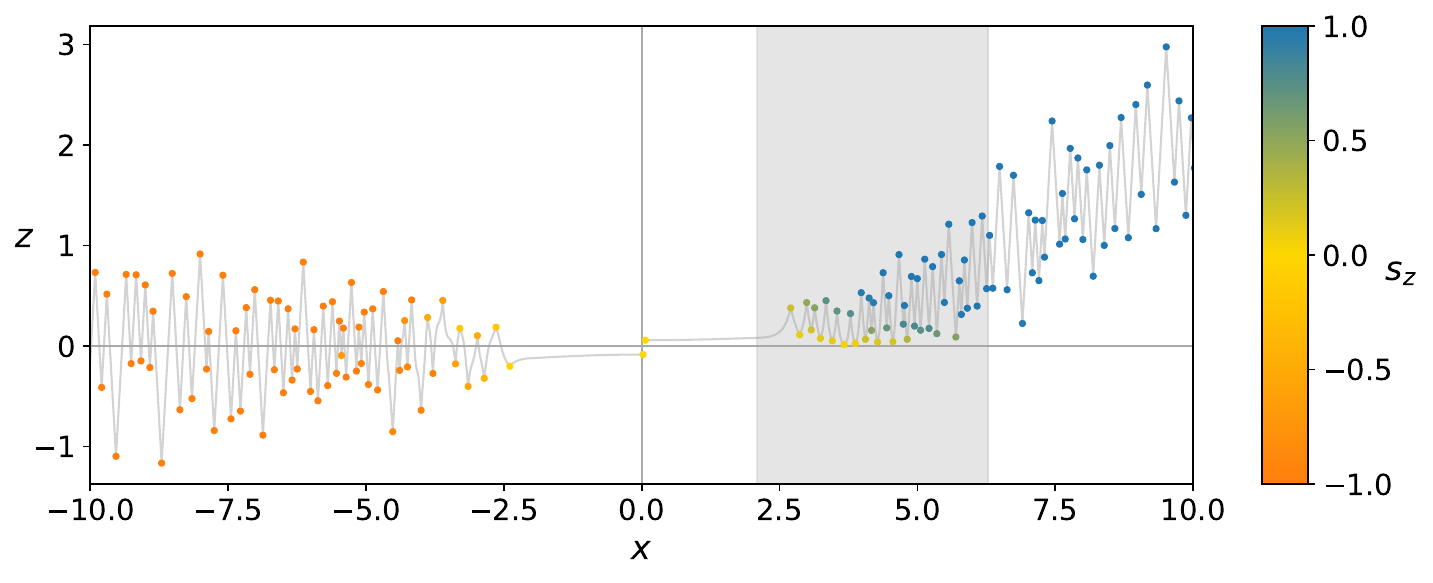}
    \caption{The same trajectories as in Fig.~\ref{fig:sg1}, with dots at the location of the chirality jumps and the color indicating the value of $s_z$. }
    \label{fig:sg2}
\end{figure}

Let us now compare this to the case where particle~1 is sent through a Stern-Gerlach device.
In this case, the spinor is
\begin{equation}
\Psi_{\textrm{SG}}({\bf x}_1,{\bf x}_2,t)=
\psi_{-p,{\bf 0},+}({\bf x}_2,t) \left[ \psi_{p,{\bf B},+}({\bf x}_1,t)  \begin{pmatrix}  1 \\ 0 \end{pmatrix}\begin{pmatrix}  0 \\ 1 \end{pmatrix}  - \psi_{p,{\bf B},-}({\bf x}_1,t)  \begin{pmatrix}  0 \\ 1 \end{pmatrix} \begin{pmatrix}  1 \\ 0 \end{pmatrix}\right].
\label{eq:entangled}
\end{equation}
The spin vectors remain opposite, {\it i.e.}, 
\be
{\bf s}_1 ({\bf x}_1, {\bf x}_2,t)= - {\bf s}_2({\bf x}_1, {\bf x}_2,t) = \frac{|\psi_{p,{\bf B},+}({\bf x}_1,t)|^2 - |\psi_{p,{\bf B},-}({\bf x}_1,t)|^2 }{|\psi_{p,{\bf B},+}({\bf x}_1,t)|^2 + |\psi_{p,{\bf B},-}({\bf x}_1,t)|^2} {\bf e}_z.
\en
Note that the spin vectors only depend on ${\bf x}_1$, so their value along a trajectory only depends on the position of particle~1.
Their direction is always along the $z$-axis.
The spin vector ${\bf s}_1$ will be pointing along the positive (negative) $z$-axis if the $z$-coordinate of particle~1 is greater (smaller) than zero.
Initially, at time $t=0$, the spin vectors are zero.
The magnetic field causes a separation of the wave functions $\psi_{p,{\bf B},+}$ and $\psi_{p,{\bf B},-}$, causing the spin vector of both particles to become nonzero.
As is shown in Figs.~\ref{fig:sg1}--\ref{fig:sg2}, once particle~1 enters the Stern-Gerlach device it starts zig-zagging and this causes particle~2 to also start zig-zagging. While particle~1 goes up in the $z$-direction, thereby displaying the result of the spin measurement, particle~2 keeps zig-zagging around $z=0$. Other trajectories display similar behavior, with particle~1 ending up either {\it up} or {\it down}, depending on the initial position and the jumps.

After passage through the magnetic field, there is an effective collapse of $\Psi_{SG}$. In the present case, this is to
\begin{equation}
\psi_{-p,{\bf 0},+}({\bf x}_2,t) \psi_{p,{\bf B},+}({\bf x}_1,t)  \begin{pmatrix}  1 \\ 0 \end{pmatrix}\begin{pmatrix}  0 \\ 1 \end{pmatrix}  ,
\label{eq:collapsed}
\end{equation}
which is a product wave function. So the dynamics of particle~1 is determined by $\psi_{p,{\bf B},+}({\bf x}_1,t) (1,0)^\tran$, while that of particle~2 is determined by $\psi_{-p,{\bf 0},+}({\bf x}_2,t) (0,1)^\tran$. If a spin measurement in the $z$-direction were to be performed on particle~2, it would yield the result spin {\it down}, in agreement with the Born rule which predicts perfect anti-correlation of the spin measurement results.

If the Stern-Gerlach device were rotated in the $y$-direction, both particles would end up zig-zagging in that direction (since the initial spinor is rotationally invariant).
So whether particle~1 is moving through a Stern-Gerlach device or not, as well as the direction of the corresponding magnetic field, have an effect on particle~2.
This effect is instantaneous and is an instance of nonlocality.
(That was also illustrated for the dynamics without the zig-zagging  in \cite{dewdney87b,dewdney88b,norsen14b}.)

\section{Conclusions}
Textbooks in quantum mechanics typically present an {\em operational} rather than a {\it mechanical} derivation of the Stern-Gerlach phenomenology. Trajectories do not appear. The Stern-Gerlach result is often {\it postulated} rather than {\it derived} from the microscopic dynamics, often serving to illustrate or motivate the axioms of standard quantum mechanics. Bohmian mechanics provides such a microscopic dynamics in terms of particle trajectories. There is no ambiguity on what is meant by ``measuring spin'', or by the ``collapse'' of the wave function. 

In the present paper, we have investigated the Stern-Gerlach setup using the Bohmian zig-zag dynamics, which is motivated by the fundamentally massless nature of particles. We assumed a Stern-Gerlach-type interaction for electrons, showing how an inhomogeneous magnetic field affects the zig-zagging. 
In the case of the EPR-pair, we showed how the Stern-Gerlach measurement of one electron also affects the motion of the other, determining the possible outcomes in the case of a spin measurement for this other particle.

While our particular modelling of the Stern-Gerlach setup is suitable for atoms, it is not physically implementable as such for electrons. However, it is expected that the qualitative behaviour will be obtained in physically realizable implementations of Stern-Gerlach-like spin polarization mechanisms.
On the other hand, it might also be interesting to try to model an actual atom in a Stern-Gerlach setup using the zig-zag dynamics. Due to its composite nature, the zig-zagging of the atom might very well disappear. Perhaps there is some internal zig-zagging of its consituent particles, but even that might be absent due to the entanglement.

Celebrating the centennial of the Stern-Gerlach experiment is obviously a tribute to the original discoverers Otto Stern and Walther Gerlach, who, first by nonbelieving and by misinterpretation, actually came to exhibit a major foundation of Nature's working on the subatomic scale.
It became a cornerstone of modern quantum mechanics.
However, we are especially happy to dedicate this paper to the memory of Detlef Dürr, who by insisting on clarity and the importance of a mechanical picture, has encouraged many of us to maintain a Boltzmannian perspective in terms of particle trajectories in the modeling and study of quantum matter as well \cite{duerr01, duerr09,duerr12}:
\begin{quote}
``We should deny ourselves from any picture of reality.''  But are man's thoughts something different from these images?  It is only God of whom we must not and cannot make any  picture.\\
--- Boltzmann's first answer to the {\it energeticists}, in: {\it A word from mathematics on energetics}, Popul\"are Schriften, Barth, Leipzig (1905) \cite{cercignani}. 
\end{quote}

\vspace{1cm}
\noindent {\bf Acknowledgment}:
WS acknowledges support from the Research Foundation Flanders (Fonds Wetenschappelijk Onderzoek, FWO), Grant No. G0C3322N. We thank Herman Batelaan, Siddhant Das and Travis Norsen for useful discussions, and the anonymous referees for helpful suggestions and comments.

\printbibliography

\end{document}